# Analytical study on the Applicability of Ultra Generalized Exponential–Hyperbolic Potential to Predict the Mass-Spectra of the Heavy Mesons

E. P. Inyang*[1], E. P. Inyang[1], J.E.Ntibi[1], E. E. Ibekwe[2], and E. S. William[1]

[1]Theoretical Physics Group, Department of Physics, University of Calabar, P.M.B 1115 Calabar Nigeria

[2]Department of Physics, Akwa Ibom State University, Ikot Akpaden, P.M.B 1167, Uyo, Nigeria

*Corresponding author email: etidophysics@gmail.com*

## Abstract

We solved the Klein-Gordon equation analytically using the Nikiforov-Uvarov method to obtain the energy eigenvalues and corresponding wavefunction in terms of Laguerre polynomials with the ultra generalized exponential –hyperbolic potential. The present results are applied for calculating the mass spectra of heavy mesons such as charmonium ($c\bar{c}$) and bottomonium ($b\bar{b}$) for different quantum states. The present potential provides excellent results in comparison with experimental data with a maximum error of $0.0059\ GeV$ and work of other researchers.

## 1.0 Introduction

The solution of the spectral problem for the Klein-Gordon equation with spherically symmetric potentials is of major concern in describing the spectra of heavy mesons. Potential models offer a rather good description of the mass spectra of quarkonium systems such as bottomonium and charmonium [1-5]. In simulating the interaction for these systems, confining-type potentials are generally used. The holding potential is the so-called Cornell potential with two terms, one of which is responsible for the Coulomb interaction of quarks and the other corresponds to the confinement of the quark [6]. Although this potential, proposed to describe quarkonia with heavy quarks, have been used for a long time, nevertheless the problem of finding the inter-quark potential with exponential type potential still remains incompletely solved. In recent times the solutions of the Schrödinger equation (SE) and Klein-Gordon equation (KGE) under the quarkonium interaction potential model such as the Cornell or the Killingbeck potentials have attracted mush interest to researchers [7-15]. The KGE with some potential can be solved exactly for $l = 0$, but insolvable for any arbitrary angular momentum quantum number $l \neq 0$. In this case, several approximate techniques are employed in obtaining the solution. For instance, such techniques include, asymptotic iteration method(AIM)[16] Laplace transformation method [17], super symmetric quantum mechanics method (SUSQM)[18-20], Nikiforov-Uvarov(NU) method [21-34],series expansion method (SEM) [35-37], analytical exact iterative method(AEIM)[38], and others [39].

Various exponential type potentials have been studied by many researchers, such as Hellmann plus Hulthen potential [40], Kratzer plus screened Coulomb potential [25], Yukawa potential [41], class of Yukawa plus Eckart potential [29] and many more. The trigonometric hyperbolic potential plays a vital role in atomic and molecular physics. Since it can be used to model inter-atomic and inter-molecular forces [42, 43].

The ultra generalized exponential –hyperbolic potential (UGEHP) takes the form [44]

$$V(r) = \frac{ae^{-4\alpha r} + be^{-2\alpha r}}{r^2} + \frac{ce^{-2\alpha r} - dCosh(\eta\alpha r)e^{-\alpha r} + gCo\sec h\alpha re^{\alpha r}}{r} + f \quad (1)$$

Where $a, b, c, d, \eta, g$ and $f$ are potential strengths and $\alpha$ is the screening parameter. When $\eta = 1$, then

$$\left. \begin{array}{l} Cosh\alpha r = \dfrac{e^{\alpha r} + e^{-\alpha r}}{2} \\ Cosech\,\alpha r = \dfrac{2}{e^{\alpha r} - e^{-\alpha r}} \end{array} \right\} \quad (2)$$

We carry out series expansion of the exponential terms in Eqs. (1) and (2) up to order three, in order to model the potential to interact in the quark-antiquark system and substitute the results into Eq.(1) this yields,

$$V(r) = \frac{\beta_0}{r^2} - \frac{\beta_1}{r} + \beta_2 r - \beta_3 r^2 + \beta_4 \quad (3)$$

Where

$$\left. \begin{array}{l} \beta_0 = a + b, \beta_1 = 4a\alpha + 2b\alpha + d - g, \beta_2 = 2c\alpha^2 + \alpha d \\ \beta_3 = \alpha(d - g), \beta_4 = 8a\alpha^2 + 2b\alpha^2 - 2c\alpha - \alpha d + g\alpha + f \end{array} \right\} \quad (4)$$

The third term of Eq.(3) is a linear term for confinement feature and the second term is the Coulomb potential that describes the short distance between quarks.

Researchers in recent times have obtained the mass spectrum of the quarkonium systems using different techniques [45-47]. For instance, Inyang et al. [45] examined heavy quarkonia characteristics in the general framework of SE with extended Cornell potential using Exact quantization rule. Furthermore, Omugbe et al.[39] obtained the heavy and heavy-light spectra in non-relativistic regime with Killingbeck potential plus an inversely quadratic potential model using the WKB method. In addition, Inyang et al.[46] obtained the Klein-Gordon equation solutions for the Yukawa potential using the Nikiforov-Uvarov method. The energy eigenvalues were obtained both in relativistic and non-relativistic regime. They results were applied to calculate heavy-meson masses. Therefore, in this present work, we aim at studying the KGE with the ultra generalized exponential –hyperbolic potential (UGEHP) using the NU method to obtain the mass spectra of heavy mesons such as charmonium $(c\bar{c})$ and bottomonium $(b\bar{b})$. To the best of our knowledge this study in not in literature. The study will be carried out in threefold. We will first model the potential to interact in the quark-antiquark system, thereafter we solved the model potential with KGE using NU method and finally the mass spectra are calculated.

**2.0 Bound state solution of the Klein-Gordon equation with the ultra generalized exponential –hyperbolic potential (UGEHP)**

The Klein-Gordon equation for a spinless particle for $\hbar = c = 1$ in N-dimensions is given as [46]

$$\left[-\nabla^2 + (M + S(r))^2 + \frac{(N + 2l - 1)(N + 2l - 3)}{4r^2}\right]\psi(r,\theta,\varphi) = [E_{nl} - V(r)]^2 \psi(r,\theta,\varphi) \qquad (5)$$

where $\nabla^2$ is the Laplacian, $M$ is the reduced mass, $E_{nl}$ is the energy spectrum, $n$, and $l$ are the radial and orbital angular momentum quantum numbers respectively. It is well known that for the wavefunction to satisfy the boundary conditions it can be rewritten as

$$\psi(r,\theta,\varphi) = \frac{R_{nl}}{r} Y_{lm}(\theta,\varphi) \qquad (6)$$

The angular component of the wavefunction could be separated leaving only the radial part as shown below

$$\frac{d^2R(r)}{dr^2} + \left[\left(E_{nl}^2 - M^2\right) + V^2(r) - S^2(r) - 2\left(E_{nl}V(r) + MS(r)\right) - \frac{(N+2l-1)(N+2l-3)}{4r^2}\right]R(r) = 0 \qquad (7)$$

Thus, for equal vector and scalar potentials $V(r) = S(r) = 2V(r)$, then Eq.(7) becomes

$$\frac{d^2R(r)}{dr^2} + \left[\left(E_{nl}^2 - M^2\right) - 2V(r)\left(E_{nl} + M\right) - \frac{(N+2l-1)(N+2l-3)}{4r^2}\right]R(r) = 0 \qquad (8)$$

Upon substituting Eq.(3) into Eq.(8), we obtain

$$\frac{d^2R(r)}{dr^2} + \left[\begin{array}{c}\left(E_{nl}^2 - M^2\right) + \left(-\frac{2\beta_0}{r^2} + \frac{2\beta_1}{r} - 2\beta_2 r + 2\beta_3 r^2 - 2\beta_4\right)\left(E_{nl} + M\right) \\ -\frac{(N+2l-1)(N+2l-3)}{4r^2}\end{array}\right]R(r) = 0 \qquad (9)$$

In order to transform the coordinate from $r$ to $x$ in Eq. (9), we set

$$x = \frac{1}{r} \qquad (10)$$

This implies that the 2$^{nd}$ derivative in Eq.(10) becomes;

$$\frac{d^2R(r)}{dr^2} = 2x^3\frac{dR(x)}{dx} + x^4\frac{d^2R(x)}{dx^2} \qquad (11)$$

Substituting Eqs.(10) and (11) into Eq.(9) we obtain

$$\frac{d^2R(x)}{dx^2} + \frac{2}{x}\frac{dR}{dx} + \frac{1}{x^4}\left[\begin{array}{c}\left(E_{nl}^2 - M^2\right) + \left(-2\beta_0 x^2 + 2\beta_1 x - \frac{2\beta_2}{x} + \frac{2\beta_3}{x^2} - 2\beta_4\right)\left(E_{nl} + M\right) \\ -\frac{(N+2l-1)(N+2l-3)x^2}{4}\end{array}\right]R(x) = 0 \qquad (12)$$

Next, we propose the following approximation scheme on the term $\frac{\beta_2}{x}$ and $\frac{\beta_3}{x^2}$.

Let us assume that there is a characteristic radius $r_0$ of the meson. Then the scheme is based on the expansion of $\frac{\beta_2}{x}$ and $\frac{\beta_3}{x^2}$ in a power series around $r_0$; i.e., around $\delta \equiv \frac{1}{r_0}$, in the x-space up to the second order. This is similar to Pekeris approximation, which helps to deform the centrifugal term such that the potential can be solved by NU method [47].

Setting $y = x - \delta$ and around $y = 0$, it can be expanded into a series of powers as;

$$\frac{\beta_2}{x} = \frac{\beta_2}{y+\delta} = \frac{\beta_2}{\delta\left(1+\frac{y}{\delta}\right)} = \frac{\beta_2}{\delta}\left(1+\frac{y}{\delta}\right)^{-1} \tag{13}$$

which yields

$$\frac{\beta_2}{x} = \beta_2\left(\frac{3}{\delta} - \frac{3x}{\delta^2} + \frac{x^2}{\delta^3}\right) \tag{14}$$

Similarly,

$$\frac{\beta_3}{x^2} = \beta_3\left(\frac{6}{\delta^2} - \frac{8x}{\delta^3} + \frac{3x^2}{\delta^4}\right) \tag{15}$$

By substituting Eqs.(14) and (15) into Eq.(12), we obtain

$$\frac{d^2 R(\mathrm{x})}{dx^2} + \frac{2x}{x^2}\frac{dR(\mathrm{x})}{dx} + \frac{1}{x^4}\left[-\varepsilon + \beta x - \gamma x^2\right]R(\mathrm{x}) = 0 \tag{16}$$

where

$$\left.\begin{aligned}
-\varepsilon &= \left((E_{nl}^2 - M^2) - \frac{6\beta_2}{\delta}(E_{nl} + M) + \frac{12\beta_3}{\delta^2}(E_{nl} + M) - 2\beta_4(E_{nl} + M)\right) \\
\beta &= \left(2\beta_1(E_{nl} + M) + \frac{6\beta_2}{\delta^2}(E_{nl} + M) - \frac{16\beta_3}{\delta^3}(E_{nl} + M)\right) \\
\gamma &= \left(2\beta_0(E_{nl} + M) + \frac{2\beta_2}{\delta^3}(E_{nl} + M) - \frac{6\beta_3}{\delta^4}(E_{nl} + M) + \frac{(N+2l-1)(N+2l-3)}{4}\right)
\end{aligned}\right\} \tag{17}$$

Comparing Eq. (16) and Eq. (A1) we obtain

$$\left.\begin{aligned}
\tilde{\tau}(\mathrm{x}) &= 2x, \ \sigma(\mathrm{x}) = x^2 \\
\tilde{\sigma}(\mathrm{x}) &= -\varepsilon + \beta x - \gamma x^2 \\
\sigma'(\mathrm{x}) &= 2x, \ \sigma''(\mathrm{x}) = 2
\end{aligned}\right\} \tag{18}$$

We substitute Eq. (18) into Eq. (A9) and obtain

$$\pi(\mathrm{x}) = \pm\sqrt{\varepsilon - \beta x + (\gamma + k)x^2} \tag{19}$$

To determine $k$, we take the discriminant of the function under the square root, which yields

$$k = \frac{\beta^2 - 4\gamma\varepsilon}{4\varepsilon} \tag{20}$$

We substitute Eq. (20) into Eq. (19) and have

$$\pi(\mathrm{x}) = \pm\left(\frac{\beta x}{2\sqrt{\varepsilon}} - \frac{\varepsilon}{\sqrt{\varepsilon}}\right) \tag{21}$$

We take the negative part of Eq. (21) and differentiate, which yields

$$\pi'_-(x) = -\frac{\beta}{2\sqrt{\varepsilon}} \tag{22}$$

By substituting Eqs. (18) and (22) into Eq. (A7) we have

$$\tau(x) = 2x - \frac{\beta x}{\sqrt{\varepsilon}} + \frac{2\varepsilon}{\sqrt{\varepsilon}} \tag{23}$$

Differentiating Eq. (23) we have

$$\tau'(x) = 2 - \frac{\beta}{\sqrt{\varepsilon}} \tag{24}$$

By using Eq. (A10), we obtain

$$\lambda = \frac{\beta^2 - 4\gamma\varepsilon}{4\varepsilon} - \frac{\beta}{2\sqrt{\varepsilon}} \tag{25}$$

And using Eq. (A11), we obtain

$$\lambda_n = \frac{n\beta}{\sqrt{\varepsilon}} - n^2 - n \tag{26}$$

Equating Eqs. (25) and (26), and substituting Eqs. (4) and (17), yields the energy eigenvalue equation of the **UGEHP** in the relativistic limit as

$$M^2 - E_{nl}^2 = \frac{6(2c\alpha^2 + \alpha d)}{\delta}(E_{nl} + M) - \frac{12\alpha(d-g)}{\delta^2}(E_{nl} + M)$$
$$+ 2(8a\alpha^2 + 2b\alpha^2 - 2c\alpha - \alpha d + g\alpha + f)(E_{nl} + M)$$
$$+ \frac{1}{4}\left[\frac{2(4a\alpha + 2b\alpha + d - g)(E_{nl} + M) + \frac{6(2c\alpha^2 + \alpha d)(E_{nl} + M)}{\delta^2} - \frac{16\alpha(d-g)(E_{nl} + M)}{\delta^3}}{n + \frac{1}{2} + \sqrt{\frac{1}{4} - 2(a+b)(E_{nl} + M) + \frac{2(2c\alpha^2 + \alpha d)}{\delta^3}(E_{nl} + M) - \frac{6\alpha(d-g)}{\delta^4}(E_{nl} + M) + \frac{(N+2l-1)(N+2l-3)}{4}}}\right]^2 \tag{27}$$

**2.1 Non relativistic limit**

In this section, we consider the non-relativistic limit of Eq.(27). Considering a transformation of the form: $M + E_{nl} \to \frac{2\mu}{\hbar^2}$ and $M - E_{nl} \to -E_{nl}$, where $\mu$ is the reduced mass, and substituting it into Eq.(27), we have the non-relativistic energy eigenvalues equation as

$$E_{nl} = \frac{12\alpha(d-g)}{\delta^2} - \frac{6(2c\alpha^2 - \alpha d)}{\delta} - 2(8a\alpha + 2b\alpha^2 - 2c\alpha - \alpha d + \alpha g + f)$$

$$-\frac{\hbar^2}{8\mu}\left[\frac{\frac{4\mu}{\hbar^2}(4a\alpha + 2b\alpha + d - g) + \frac{12\mu}{\delta^2\hbar^2}(2c\alpha^2 + \alpha d) - \frac{32\mu\alpha}{\delta^3\hbar^2}(d-g)}{n + \frac{1}{2} + \sqrt{\frac{1}{4} - \frac{4\mu}{\hbar^2}(a+b) + \frac{4\mu}{\delta^3\hbar^2}(2c\alpha^2 + \alpha d) - \frac{12\mu\alpha}{\delta^4\hbar^2}(d-g) + \frac{(N+2l-1)(N+2l-3)}{4}}}\right]^2 \quad (28)$$

The unnarmalized wavefunction in terms of Laguerre polynomials is given as

$$\psi(s) = B_{nl} s^{-\frac{\alpha_1}{2\sqrt{\varepsilon}}} e^{-\frac{\varepsilon}{s\sqrt{\varepsilon}}} L_n^{\frac{\alpha_1}{\sqrt{\varepsilon}}}\left(\frac{2\varepsilon}{s\sqrt{\varepsilon}}\right), \quad (29)$$

where $L_n$ is the associated Laguerre polynomials and $B_{nl}$ is normalization constant, which can be obtain from

$$\int_0^\infty |B_{nl}(r)|^2 dr = 1 \quad (30)$$

## 3.0 Results and discussion

### 3.1 Results

We calculate mass spectra of the heavy quarkonium system such as charmonium and bottomonium in 3-dimensional space ($N = 3$) that have the quark and antiquark flavor, and apply the following relation [48]

$$M = 2m + E_{nl}^{N=3}, \quad (31)$$

where $m$ is quarkonium bare mass, and $E_{nl}^{N=3}$ is energy eigenvalues. By substituting Eq. (28) into Eq. (31) we obtain the mass spectra for UGEHP as:

$$M = 2m + \frac{12\alpha(d-g)}{\delta^2} - \frac{6(2c\alpha^2 - \alpha d)}{\delta} - 2(8a\alpha + 2b\alpha^2 - 2c\alpha - \alpha d + \alpha g + f)$$

$$-\frac{\hbar^2}{8\mu}\left[\frac{\frac{4\mu}{\hbar^2}(4a\alpha + 2b\alpha + d - g) + \frac{12\mu}{\delta^2\hbar^2}(2c\alpha^2 + \alpha d) - \frac{32\mu\alpha}{\delta^3\hbar^2}(d-g)}{n + \frac{1}{2} + \sqrt{\frac{1}{4} - \frac{4\mu}{\hbar^2}(a+b) + \frac{4\mu}{\delta^3\hbar^2}(2c\alpha^2 + \alpha d) - \frac{12\mu\alpha}{\delta^4\hbar^2}(d-g) + \frac{(N+2l-1)(N+2l-3)}{4}}}\right]^2 \quad (32)$$

**Table 1**: Mass spectra of charmonium in $(GeV)$

$$\begin{pmatrix} a = -22.17885\ GeV, b = 13.73217\ GeV, c = 10.73524\ GeV, d = 3.010241\ GeV, g = 10.64213\ GeV, \\ f = 0.05\ GeV, \alpha = 0.01, \delta = 1.00252\ GeV, m_c = 1.209\ GeV, N = 3, \hbar = 1, \mu = 0.6045\ GeV \end{pmatrix}$$

| State | Present work | [7] | [48] | Experiment[49] |
|---|---|---|---|---|
| 1S | 3.096 | 3.096 | 3.096 | 3.096 |
| 2S | 3.686 | 3.686 | 3.672 | 3.686 |
| 1P | 3.526 | 3.255 | 3.521 | 3.525 |
| 2P | 3.767 | 3.779 | 3.951 | 3.773 |
| 3S | 4.040 | 4.040 | 4.085 | 4.040 |
| 4S | 4.262 | 4.269 | 4.433 | 4.263 |
| 1D | 3.768 | 3.504 | 3.800 | 3.770 |
| 2D | 4.034 | - | - | 4.159 |
| 1F | 4.162 | - | - | - |

**Table 2**: Mass spectra of bottomonium in $(GeV)$

$$\begin{pmatrix} a = -20.99857 \text{ } GeV, b = 13.6254385 \text{ } GeV, c = 13.73524 \text{ } GeV, d = 4.110240 \text{ } GeV, g = 11.542130 \text{ } GeV, \\ f = 0.05 \text{ } GeV, \alpha = 0.01, \delta = 1.00252 \text{ } GeV, m_c = 4.823 \text{ } GeV, N = 3, \hbar = 1, \mu = 2.4115 \text{ } GeV \end{pmatrix}$$

| State | Present work | [7] | [48] | Experiment[49] |
|---|---|---|---|---|
| 1S | 9.460 | 9.460 | 9.4620 | 9.460 |
| 2S | 10.023 | 10.023 | 10.027 | 10.023 |
| 1P | 9.761 | 9.619 | 9.9630 | 9.899 |
| 2P | 10.261 | 10.114 | 10.299 | 10.260 |
| 3S | 10.355 | 10.355 | 10.361 | 10.355 |
| 4S | 10.579 | 10.567 | 10.624 | 10.580 |
| 1D | 9.998 | 9.864 | 10.209 | 10.164 |
| 2D | 10.206 | - | - | - |
| 1F | 10.109 | - | - | - |

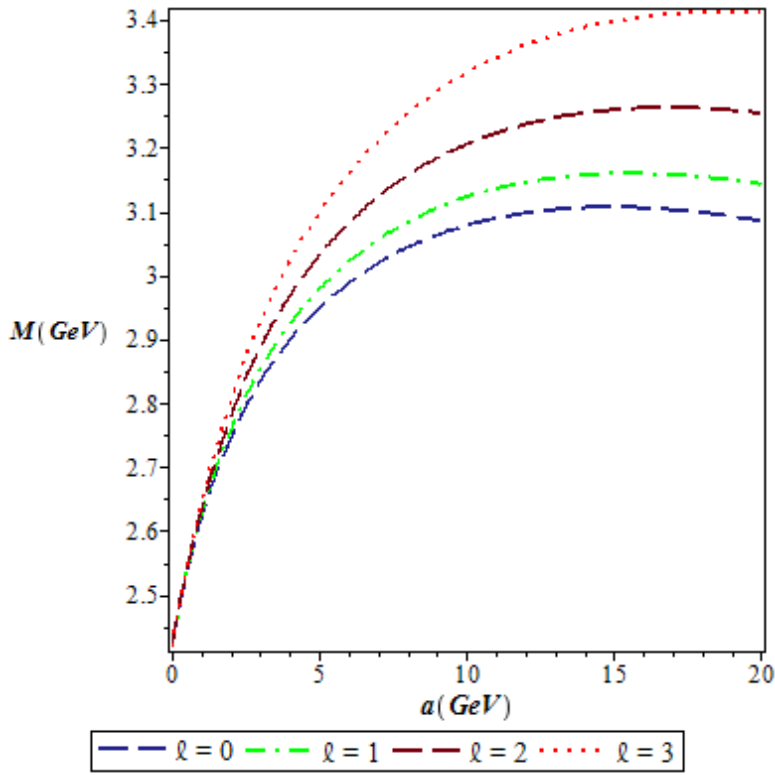

Figure 1: Variation of mass spectra with potential strength $(a)$ for different quantum numbers

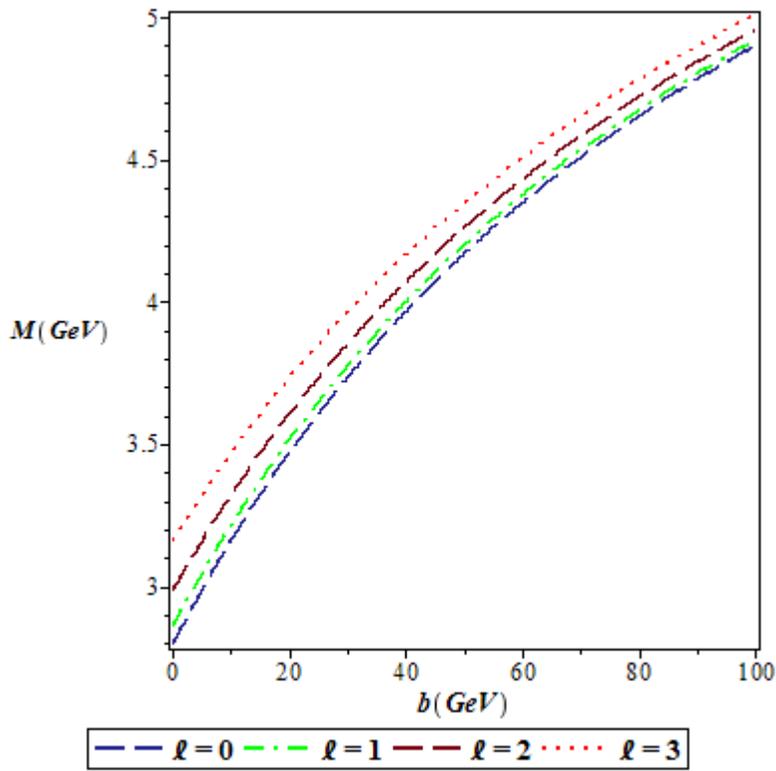

Figure 2: Variation of mass spectra with potential strength $(b)$ for different quantum numbers

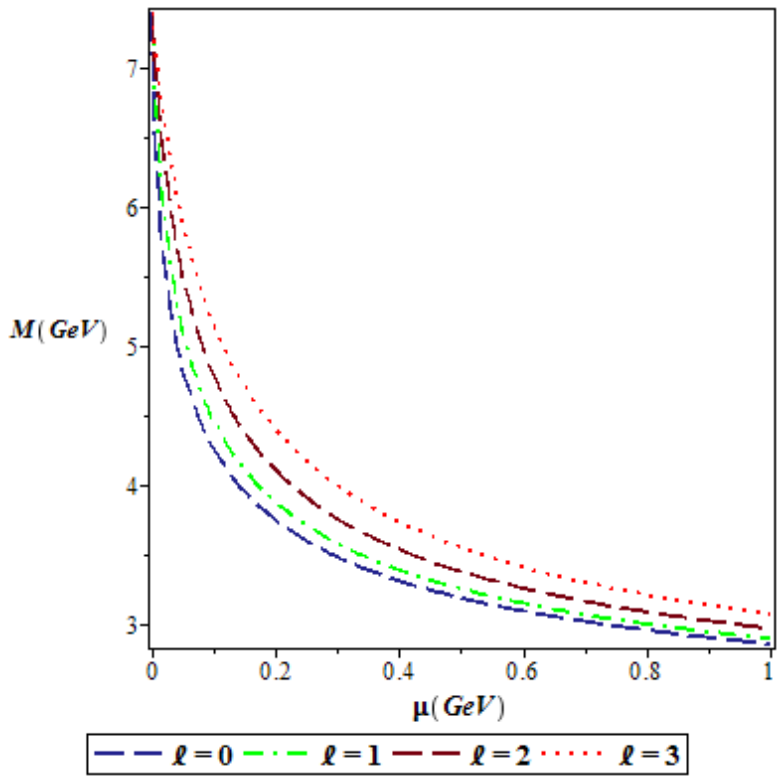

Figure 3: Variation of mass spectra with reduced mass $(\mu)$ for different quantum numbers

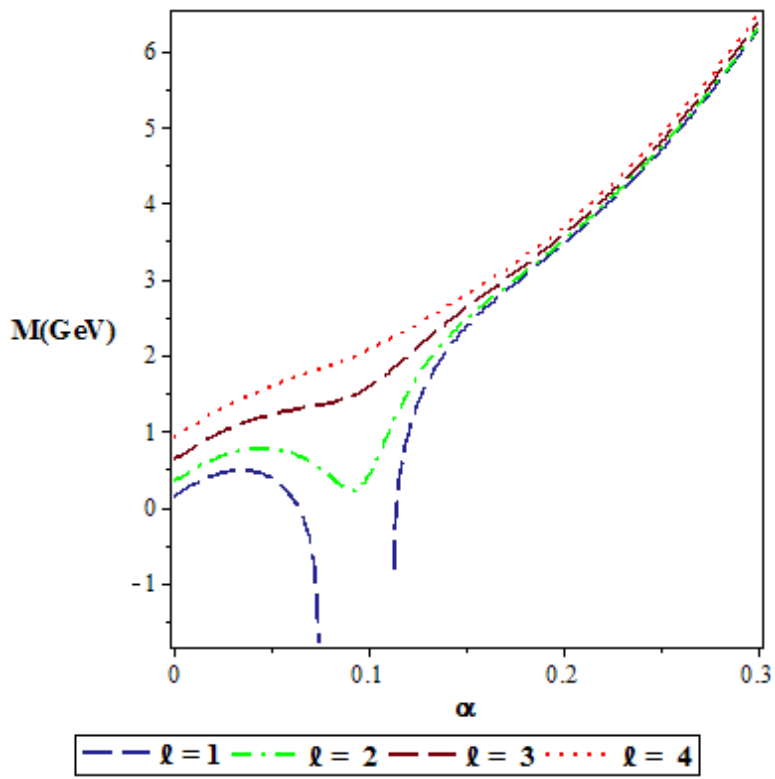

Figure 4: Variation of mass spectra with screening parameter $(\alpha)$ for different quantum numbers

## 3.2 Discussion of results

We calculate mass spectra of charmonium and bottomonium for states from 1S, 2S, 1P, 2P, 3S, 4S, 1D, 2D, and 1F, by using Eq. (32). The potential parameters of Eq. (32) were then obtained by solving two algebraic equations in the case of charmonium and bottomonium, respectively.

The experimental data were taken from [49]. For bottomonium $b\bar{b}$ and charmonium $c\bar{c}$ systems we adopt the numerical values of these masses as $m_b = 4.823\,GeV$ and $m_c = 1.209\,GeV$ [50]. Then, the corresponding reduced mass are $\mu_b = 2.4115\,GeV$ and $\mu_c = 0.6045\,GeV$, respectively. We note that calculation of mass spectra of charmonium and bottomonium are in good agreement with experimental data and work of other researchers, in Refs.[7, 48] as presented in Tables 1 and 2. In order to test for the accuracy of the predicted results determined numerically, we used a Chi squared function to determine the error between the experimental data and theoretical predicted values. The maximum error in comparison with the experimental data is found to be $0.0059\,GeV$. We plotted the variation of mass spectra energy with respect to potential strengths, reduced mass $(\mu)$ and screening parameter $(\alpha)$ respectively. In Figs. 1 and 2, the mass spectra energy increases as the potential strength increases for different quantum number. In Fig. 3, it is observed that the mass spectra energy decreases exponentially as the reduced mass increases for various angular quantum numbers. Finally, increase in mass spectra energy, is observed as the screening parameter increases.

## 4.0 Conclusion

In this study, we model the adopted ultra generalized exponential–hyperbolic potential to interact in quark-antiquark system. We obtained the approximate solutions of the Schrödinger equation for energy eigenvalues and unnormalized wave function using the NU method. We applied the present results to compute heavy-meson masses of charmonium and bottomonium for different quantum states. The result agreed with experimental data with a maximum error of $0.0059\,GeV$ and work of other researchers. Mass spectra variation with potential strengths, reduced mass $(\mu)$ and screening parameter $(\alpha)$ were plotted and discussed.

## APPENDIX A: Review of Nikiforov-Uvarov(NU) method

The NU method was proposed by Nikiforov and Uvarov [51] to transform Schrödinger-like equations into a second-order differential equation via a coordinate transformation $x = x(r)$, of the form

$$\psi''(x) + \frac{\tilde{\tau}(x)}{\sigma(x)}\psi'(s) + \frac{\tilde{\sigma}(x)}{\sigma^2(x)}\psi(x) = 0 \tag{A1}$$

where $\tilde{\sigma}(x)$, and $\sigma(x)$ are polynomials, at most second degree and $\tilde{\tau}(x)$ is a first-degree polynomial. The exact solution of Eq.(A1) can be obtain by using the transformation.

$$\psi(x) = \phi(x)y(x) \tag{A2}$$

This transformation reduces Eq.(A1) into a hypergeometric-type equation of the form

$$\sigma(x)y''(x) + \tau(x)y'(x) + \lambda y(x) = 0 \tag{A3}$$

The function $\phi(x)$ can be defined as the logarithm derivative

$$\frac{\phi'(x)}{\phi(x)} = \frac{\pi(x)}{\sigma(x)} \tag{A4}$$

With $\pi(x)$ being at most a first-degree polynomial. The second part of $\psi(x)$ being $y(x)$ in Eq.(A2) is the hypergeometric function with its polynomial solution given by Rodrigues relation as

$$y(x) = \frac{B_{nl}}{\rho(x)}\frac{d^n}{dx^n}\left[\sigma^n(x)\rho(x)\right] \tag{A5}$$

Where $B_{nl}$ is the normalization constant and $\rho(x)$ the weight function which satisfies the condition below;

$$(\sigma(x)\rho(x))' = \tau(x)\rho(x) \tag{A6}$$

where also

$$\tau(x) = \tilde{\tau}(x) + 2\pi(x) \tag{A7}$$

For bound solutions, it is required that

$$\tau'(x) < 0 \tag{A8}$$

The eigenfunctions and eigenvalues can be obtained using the definition of the following function $\pi(x)$ and parameter $\lambda$, respectively:

$$\pi(x) = \frac{\sigma'(x) - \tilde{\tau}(x)}{2} \pm \sqrt{\left(\frac{\sigma'(x) - \tilde{\tau}(x)}{2}\right)^2 - \tilde{\sigma}(x) + k\sigma(x)} \tag{A9}$$

and

$$\lambda = k_- + \pi'_-(x) \tag{A10}$$

The value of $k$ can be obtained by setting the discriminant in the square root in Eq. (A9) equal to zero. As such, the new eigenvalues equation can be given as

$$\lambda + n\tau'(x) + \frac{n(n-1)}{2}\sigma''(x) = 0, (n = 0,1,2,...) \tag{A11}$$